\documentclass[prl,letterpaper,preprint,showpacs]{revtex4}
\usepackage{amsmath,amssymb,bm}
\usepackage{psfrag,graphicx}
\newcommand{\vect}[1]{\bm{#1}}
\newcommand{\norm}[1]{|#1|}
\newcommand*{\B}{\beta}
\newcommand*{\GSL}{\lim_{\B\to\infty}}
\newcommand*{\Q}{Q}
\newcommand*{\half}{\frac{1}{2}}
\newcommand*{\smp}{\epsilon}
\newcommand*{\ord}[1]{\mathcal{O}(#1)}
\newcommand{\srv}[1]{\bar{#1}}
\newcommand*{\V}{V}
\newcommand*{\C}{C}
\newcommand*{\W}{W}
\newcommand*{\cp}{\mu}
\newcommand*{\epa}{\alpha}
\newcommand*{\epb}{\beta}
\newcommand*{\epc}{\lambda}
\newcommand*{\bbR}{\mathbb{R}}
\newcommand*{\EQ}{E}
\newcommand*{\EH}{E_\textsc{h}}
\newcommand*{\EXC}{E_\textsc{x}}

\newcommand*{\ECRa}{E_{\textsc{c};1}}
\newcommand*{\ECRb}{E_{\textsc{c};2}}
\newcommand*{\sEQ}{\srv{E}}
\newcommand*{\sECR}{\srv{E}_\textsc{c}}
\newcommand*{\sECRa}{\srv{E}_{\textsc{c};1}}
\newcommand*{\sECRb}{\srv{E}_{\textsc{c};2}}
\newcommand*{\sEH}{\srv{E}_\textsc{h}}
\newcommand*{\sEHF}{\srv{E}_\textsc{hf}}
\newcommand*{\sEHFsm}{\srv{E}_{\textsc{hf};\text{s}}}
\newcommand*{\sEHFos}{\srv{E}_{\textsc{hf};\text{osc}}}
\newcommand*{\sEXC}{\srv{E}_\textsc{x}}
\newcommand*{\PP}{P}

\newcommand*{\dos}{d}
\newcommand*{\dosos}{\dos_\text{osc}}
\newcommand*{\dossm}{\dos_\text{s}}
\newcommand*{\idos}{D}

\newcommand*{\deno}{\rho_0}
\newcommand*{\nf}{n}
\newcommand*{\pv}{\vect{p}}
\newcommand*{\xv}{\vect{x}}
\newcommand*{\yv}{\vect{y}}

\newcommand*{\heaviside}{\vartheta}
\newcommand*{\smhs}[1]{\heaviside(#1)}
\newcommand*{\hs}[1]{\heaviside\big(#1\big)}
\newcommand*{\kron}{\delta}

\newcommand*{\ks}[1]{\kron\big(#1\big)}
\newcommand*{\inv}{^{-1}}
\newcommand*{\xa}{\xv}
\newcommand*{\xb}{\yv}
\newcommand*{\ta}{\tau_x}
\newcommand*{\tb}{\tau_y}
\newcommand*{\FX}{(\xv)}
\newcommand*{\FY}{(\yv)}
\newcommand*{\FF}[2]{(#1|#2)}
\newcommand*{\FXY}{\FF\xv\yv}

\newcommand*{\lap}{\Delta}
\newcommand*{\isp}{\,}
\renewcommand*{\d}[1]{d #1}
\newcommand*{\D}[1]{d #1\isp}
\newcommand{\inta}[1]{\int\D{#1}}

\DeclareMathOperator{\Trace}{Tr}
\DeclareMathOperator{\trace}{tr}

\newcommand{\gAvg}[2]{\Bigl<#2\Bigr>_{#1}}
\newcommand{\scalth}[3]{\langle#1|#2|#3\rangle}
\begin{document}
%
%
%
%
%
\title{%
    Ground state energy of large atoms and quantum dots%
    }
\author{Herv\'e Kunz}
\email{herve.kunz@epfl.ch}
\author{Rico Rueedi}
\email{rico.rueedi@epfl.ch}
\affiliation{Institute of Theoretical Physics, Ecole Polytechnique F\'ed\'erale de Lausanne (EPFL), CH-1015 Lausanne, Switzerland}
\begin{abstract}
We determine the ground state energy of atoms and quantum dots whose number $N$ of electrons is large. We show that the dominant terms of the energy are those given by a semiclassical Hartree-Fock theory. Correlation effects appear at the order $N\ln N$ for atoms and the order $N$ for quantum dots. We compute them. The semiclassical Hartree-Fock theory creates oscillations in the ground state energy as a function of $N$. These oscillations reflect the dynamics of a classical particle moving in the presence of the Thomas-Fermi potential. The dynamics is regular for atoms and some dots, but we present the case of a dot where this dynamics is fully chaotic and we compute the oscillating part of the ground state energy in this case.
\end{abstract}
\maketitle
%
%
%
%
%
Since the introduction by Thomas \cite{Thomas-PCPS23} and Fermi \cite{Fermi-RANL6} of the statistical theory of atoms, many efforts have been made to systematically improve on it. For neutral atoms with $N$ electrons, an expansion of the ground state energy, whose first term is Thomas-Fermi (TF) theory, was determined up to the order $N^{5/3}$ \cite{Morgan-1996}. This expansion was based on Hartree-Fock (HF) theory. But after Lieb and Simon \cite{LiebSimon-AdvMath23} proved that TF theory becomes exact when $N\to\infty$, part of the expansion was put on a rigorous basis \cite{FeffermanSeco-HPA71}. A crucial step further, inside HF theory, was made by Schwinger and Englert \cite{EnglertSchwinger-PRA32b} who showed that there exist oscillating corrections to the ground state energy, of order $N^{4/3}$, and period $N^{1/3}$. They interpreted such terms as indicating shell effects.

We consider quantum dots as artificial atoms made of $N$ electrons, confined by a potential and constrained to move in two dimensions. The determination of the dependence of the ground state energy of quantum dots as a function of $N$ is of great interest. It was proven by Lieb, Solovej, and Yngvason \cite{LiebSolovejYngvason-PRB51} that in this case also, TF theory becomes exact when $N\to\infty$. This suggests that like in the case of atoms, corrections to it should be determined. An important difference with atoms is the fact that the confining potential isn't necessarily rotationally symmetric, and can be changed.

We have found a general formula for the ground state energy of both atoms and dots, appropriate to the case when $N\inv$ is small. It is based on the fact that in this case, we are confronted to a semiclassical mean field theory corrected by fluctuation effects. In this way, we have shown that HF theory is valid, up to a certain order in $N\inv$ where correlation effects become important. We have determined the dominant correlation correction, which in the case of atoms, is $N\ln N$.

We show that the oscillating contributions to the energy appearing in HF theory can be interpreted as coming from periodic orbits of a classical particle moving in the TF potential. In atoms, this motion is integrable. But in dots, the confining potential can be chosen so that the motion is fully chaotic, and we compute the ground state energy for a specific example of such a chaotic quantum dot.

%
%
%
%
%
We can put non-relativistic atoms and quantum dots on the same footing, if we describe them, in appropriate units, by the $d$-dimensional Hamiltonian of $N$ electrons
\begin{equation}
\srv H=
-\sum_{j=1}^N\srv\lap_j
+Z\sum_{j=1}^N\V(\srv\xv_j)
+\half\sum_{j\neq k}^N
\frac{1}{\norm{\srv\xv_j-\srv\xv_k}}.
\label{def:H:srv}
\end{equation}
For atoms ($d=3$), $Z$ is the number of protons and $\V(\srv\xv)=-\smash{\norm{\srv\xv}\inv}$. For dots ($d=2$), $Z$ will be identified with $N$, a possible dependence of $\V(\srv\xv)$ on $N$ being neglected at this stage.

For the purpose of considering systems where $Z$ (and $N$) is large, we rescale the coordinates so that $\srv H=Z^{2-2/d}H$, where
\begin{equation}
H =
-\smp^2\sum_{j=1}^N\lap_j
+\sum_{j=1}^N\V(\xv_j)
+\frac{1}{2Z}\sum_{j\neq k}^N
\frac{1}{\norm{\xv_j-\xv_k}},
\label{def:H}
\end{equation}
the parameter $\smp$, playing the role of $\hbar$, being $\smp\doteq Z^{-1/d}$. In this form, the Coulomb interaction is of mean field type, and, when $Z$ is large, the Hamiltonian has to be considered in the semiclassical limit. Our strategy to determine the ground state energy $E$ of $H$ is then the following: considering first $\smp$ and $Z$ as independent, we compute an asymptotic expansion in $Z\inv$ of the pressure $\PP$ of the Hamiltonian $H$, then take the zero temperature limit $(\B\to\infty$), and finally look at the semiclassical limit $\smp\to0$. 

With this strategy in mind we first make a Hubbard-Stratonovich transformation on the Coulomb interaction term in the coherent state path integral representation of the partition function $\Q$ \cite{NegeleOrland-1988}. After an appropriate shift, the partition function can be written as the average value over an auxiliary Gaussian field $\theta$
\begin{equation}
\Q=\gAvg{\theta}{e^{ZS(\theta)}},
\end{equation}
$S(\theta)$ being an action such that its saddle point, which dominates $\Q$ when $Z$ is large, corresponds to the Hartree approximation.

%
%
%
%
%
The final exact result for $\EQ$ can be put in the form
\begin{equation}
\EQ=\EH+\EXC+\ECRa+\ECRb.
\end{equation}
The first term is the Hartree energy, given by
\begin{equation}
\EH\doteq
\cp N-\frac{Z}{2}\int
\frac{\d\xv\d\yv}{\norm{\xv-\yv}}
\deno\FX\deno\FY
-2\int^\cp\D{e}\idos(e).
\end{equation}
$\idos(e)=\trace\smhs{e-h}$, with $\heaviside$ the Heaviside step function, is the integrated density of states of the one-body Hamiltonian $h\doteq-\smp^2\lap+\W\FX$, with $\W\FX$ the self-consistent potential given by
\begin{equation}
\W\FX\doteq
\V\FX+
\int\frac{\d\yv}{\norm{\xv-\yv}}\deno\FY.
\end{equation}
$\deno\FX$ is the density $\deno\FX\doteq 2\nf\FF\xv\xv/Z$ , where $\nf\FXY \doteq \scalth{\xv}{\smhs{\cp-h}}{\yv}$ is the density matrix associated to $h$. Finally, the chemical potential $\cp$ is fixed by the constraint $\partial_\cp\EQ=0$. The second term is the exchange energy
\begin{equation}
\EXC\doteq
-\frac{1}{Z}\int\frac{\d\xv\d\yv}{\norm{\xv-\yv}}
\nf^2\FXY.
\end{equation}
This term also appears in HF theory, where it is the dominant exchange contribution when $Z$ is large \cite{Englert-1988}. The remaining terms describe correlation effects, with
\begin{equation}
\ECRa\doteq
\GSL
-\frac{1}{2\B}\sum_{n=2}^{\infty}
\frac{1}{nZ^n}
\Trace(C\Gamma)^n,
\label{def:ECR1}
\end{equation}
and
\begin{equation}
\ECRb\doteq
\GSL
-\frac{1}{\B}
\ln
\gAvg{\theta}{\exp\Big[
-2\sum_{n=3}^{\infty}\frac{(-i)^n}{nZ^{n/2}}
\Trace(K\theta)^n\Big]}.
\end{equation}
$\C$ is the operator of kernel
\begin{equation}
\C\FF{\xv,\ta}{\xb,\tb}=
\frac{1}{\norm{\xa-\xb}}\ks{\ta-\tb},
\end{equation}
$0\leq\ta,\tb\leq\B$ being imaginary times. $K$ is the operator
\begin{equation}
K=
[\partial_\tau-\cp+h]\inv,
\end{equation}
$\Gamma$ is the operator of kernel
\begin{equation}
\Gamma\FF{\xa,\ta}{\xb,\tb}=
K\FF{\xa,\ta}{\xb,\tb}
K\FF{\xb,\tb}{\xa,\ta},
\end{equation}
and finally, $\theta(\xv,\tau)$ is a Gaussian field of covariance
$[\C\inv-\Gamma/Z\big]\FF{\xa,\ta}{\xb,\tb}$.

We recover TF theory by taking the semiclassical limit in the Hartree energy. Indeed, in this limit, the density $\deno\FX$ and the integrated density of states $D(e)$ become
\begin{equation}
\deno\FX=
\gamma_d\cp_+^{d/2}\FX,
\end{equation}
where $\cp_+\FX\doteq(\cp-\W\FX)\smhs{\cp-\W\FX}$, and
\begin{equation}
\idos(e)=
\gamma_d\inta\xv
\big(e-\W\FX\big)_+^{d/2},
\end{equation}
where $\gamma_2=(2\pi)\inv$ and $\gamma_3=(3\pi^2)\inv$, and $\cp$ is fixed by the constraint
\begin{equation}
\inta\xv\deno\FX=1.
\end{equation}

At the next stage, we compute the correlation energies. For $\ECRa$, we simply need to take the limit $\smp\to0$. For $\ECRb$, we first compute the terms up to the order $Z^{-3}$, then take the semiclassical limit. We recall that the true ground state energy is given by $\sEQ=Z^{2-2/d}E$.

In the case of neutral atoms, and moving to standard atomic units, we find
\begin{equation}
\sECRa=
-0.062\ N\ln N-
[0.034+2x_2]N+
\ord{N^{2/3}},
\end{equation}
where
\begin{equation}
x_2\doteq
\frac{1}{2}
\sum_{n=3}^\infty
\frac{1}{n}
\int_{-\infty}^\infty\frac{\d\Omega}{2\pi}\trace T_\Omega^n,
\end{equation}
$T_\Omega$ being certain operator on $L^2(\bbR^3)$, and
\begin{equation}
\sECRb=0.98\ N+\ord{N^{2/3}}.
\end{equation}
It is remarkable that the dominant term for the correlation energy $-0.062\ N\ln N$ is the same as the logarithmically divergent second order correlation energy of the jellium \cite{Gell-MannBrueckner-PR106}, an effective cut-off $\smp$ on some momentum making the integrals finite in the case of atoms. Other parts of the correlation energy are the same as in the jellium. For its correlation energy, the atoms looks like a jellium with a \emph{local} Fermi surface defined by $\pv^2\leq\cp_+\FX$.

\begin{figure}[t]
\centering
\begin{small}\begin{psfrags}
\def\PFGstripminus-#1{#1}%
\def\PFGshift(#1,#2)#3{\raisebox{#2}[\height][\depth]{\hbox{%
  \ifdim#1<0pt\kern#1 #3\kern\PFGstripminus#1\else\kern#1 #3\kern-#1\fi}}}%
\providecommand{\PFGstyle}{}%
\psfrag{CellTextDa}[cc][cc]{\PFGstyle $-\bar{E}_\textsc{cr}/N$ (Ha)}%
\psfrag{ext}[cl][cl]{\PFGstyle \footnotesize{exp}}%
\psfrag{extHF}[cl][cl]{\PFGstyle \footnotesize{ext HF}}%
\psfrag{FormBoxNnT}[cc][cc]{\PFGstyle $N$}%
\psfrag{S0}[tc][tc]{\PFGstyle \footnotesize{$ 0$}}%
\psfrag{S10}[tc][tc]{\PFGstyle \footnotesize{$10$}}%
\psfrag{S20}[tc][tc]{\PFGstyle \footnotesize{$20$}}%
\psfrag{S30}[tc][tc]{\PFGstyle \footnotesize{$30$}}%
\psfrag{S40}[tc][tc]{\PFGstyle \footnotesize{$40$}}%
\psfrag{S50}[tc][tc]{\PFGstyle \footnotesize{$50$}}%
\psfrag{S60}[tc][tc]{\PFGstyle \footnotesize{$60$}}%
\psfrag{W000}[cr][cr]{\PFGstyle \footnotesize{$0.00$}}%
\psfrag{W001}[cr][cr]{\PFGstyle \footnotesize{$0.01$}}%
\psfrag{W002}[cr][cr]{\PFGstyle \footnotesize{$0.02$}}%
\psfrag{W003}[cr][cr]{\PFGstyle \footnotesize{$0.03$}}%
\psfrag{W004}[cr][cr]{\PFGstyle \footnotesize{$0.04$}}%
\psfrag{W005}[cr][cr]{\PFGstyle \footnotesize{$0.05$}}%
\psfrag{W006}[cr][cr]{\PFGstyle \footnotesize{$0.06$}}%
\psfrag{W007}[cr][cr]{\PFGstyle \footnotesize{$0.07$}}%
\includegraphics[width=68mm]{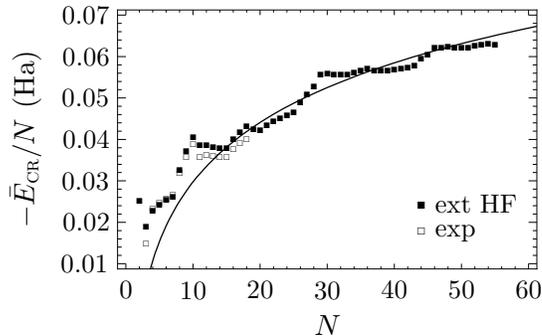}
\end{psfrags}\end{small}
\caption{Per electron correlation energy of neutral atoms. Data points correspond to experimental values \cite{ChakravortyDavidson-JPC100} and numerical values obtained with an extended HF approach \cite{ClementiCorongiu-IJQC62}, available for atoms containing up to $18$ and $55$ electrons, respectively. For sufficiently large atoms, the contribution $\sECR/N=-0.062\ln N-0.18$ deviates from experimental and extended HF values essentially by an oscillating term.}
\label{A:CR}
\end{figure}
To summarize, for neutral atoms we have found the correlation contribution $\sECR\doteq\sECRa+\sECRb$ to the ground state energy
\begin{equation}
\sECR=-0.062\ N\ln N+cN+\ord{N^{2/3}},
\label{hECR}
\end{equation}
and not, as has been suggested, a $N^{4/3}$ contribution \cite{Englert-1988,ClementiCorongiu-IJQC62}. In figure \ref{A:CR}, a comparison is made between experimental \cite{ChakravortyDavidson-JPC100} and extended Hartree-Fock \cite{ClementiCorongiu-IJQC62} values for the correlation energy, and formula \eqref{hECR} taking $c=-0.018$ for the best fit. We have not computed numerically $x_2$.

In the case of quantum dots, and in the units of \eqref{def:H:srv}, we find
\begin{equation}
\sECRa=-0.15\ N+\ord{N^{1/2}\ln N},
\end{equation}
and
\begin{equation}
\sECRb=
-0.15\ N
+ 2c_1N
+\ord{N^{1/2}},
\end{equation}
where
\begin{equation}
c_1\doteq
\frac{1}{2\pi^4}
\int\frac{\d\xv\d\yv}{\norm{\xv-\yv}}
\cp^{1/2}_+\FX
\cp^{1/2}_+\FY.
\end{equation}
It can be noticed that the dominant contribution to $\ECRa$ is universal, \emph{i.e.} independent of the confining potential.

Let us now look at the Hartree-Fock type contribution $\sEHF=\sEH+\sEXC$, which needs to be computed semiclassically. $\sEH$ depends essentially on the density of states $\dos(e)=\partial_e\idos(e)$. It is standard now to decompose $\dos(e)$ into two parts, semiclassically, as $\dos(e)=\dossm(e)+\dosos(e)$ \cite{BrackBhaduri-1997}; and, accordingly, to decompose $\sEHF$ as $\sEHF=\sEHFsm+\sEHFos$. The smooth part $\dossm(e)$ is given by an asymptotic expansion in $\smp$, whose coefficients are some integrals depending on $\W\FX$. But the oscillating part
\begin{equation}
\dosos(e)=
\sum_{\gamma}A_\smp(e,\gamma)\cos\Big(\frac{1}{\smp}S(e,\gamma)+\sigma_\gamma\frac{\pi}{2}\Big)
\end{equation}
is given by a sum over the periodic orbits $\gamma$ of a classical particle moving in the potential $\W\FX$, where $S(e,\gamma)$ is the classical action along the orbit, $\sigma_\gamma$ is the orbit's Maslov index, and $A_\smp(e,\gamma)$ depends on the orbit's period and stability. $\dosos(e)$ therefore depends crucially on the nature of the dynamics associated to $\W\FX$.

In the case of atoms, the smooth part is known up to the order $\smp\inv$, and further corrections both to the Hartree and exchange energies would be required to reach the order $\smp$ of the correlation energy. The semiclassical determination of these corrections is 
particularly delicate due to the Coulomb singularity of $\W\FX$ at the origin, and the slow decay of the Coulomb potential appearing in the exchange term. Therefore, we have not undertaken this task. The oscillating part was computed by Schwinger and Englert, who did not give a dynamical interpretation of it. The spherically symmetric potential $\W\FX$ makes the dynamics integrable in this case.

In the case of dots, we have completely determined the smooth part of $\sEHF$ at the same order as the correlation energy. It is given by
\begin{multline}
\sEHFsm=
\cp N^2
-\frac{N^2}{4\pi}
\inta{\xv}\cp_+\FX
-\frac{N^2}{8\pi^2}
\int\frac{\d\xv\d\yv}{\norm{\xv-\yv}}
\cp_+\FX\cp_+\FY
\\
-\frac{2N^{3/2}}{3\pi^3}
\inta{\xv}
\cp_+^{3/2}\FX
+\frac{N}{24\pi}
\inta{\xv}\lap\W\FX\hs{\cp\FX}+c_2N,
\end{multline}
where
\begin{equation}
c_2\doteq
\frac{1}{\pi^3}
\frac{\displaystyle
\Big[
\inta\xv\cp_+^{1/2}\FX a\FX
\Big]^2
}{\displaystyle
\inta\xv\hs{\cp\FX} a\FX},
\end{equation}
$a\FX$ being the solution of the integral equation
\begin{equation}
a\FX=
1-\frac{1}{2\pi}
\int\frac{\d\yv}{\norm{\xv-\yv}}
\hs{\cp\FY}a\FY;
\end{equation}
and where
\begin{equation}
\W\FX=
\V\FX+
\frac{1}{2\pi}
\int\frac{\d\yv}{\norm{\xv-\yv}}\cp_+\FY.
\end{equation}
\begin{figure}[t]
\centering
\begin{small}\begin{psfrags}
\def\PFGstripminus-#1{#1}%
\def\PFGshift(#1,#2)#3{\raisebox{#2}[\height][\depth]{\hbox{%
  \ifdim#1<0pt\kern#1 #3\kern\PFGstripminus#1\else\kern#1 #3\kern-#1\fi}}}%
\providecommand{\PFGstyle}{}%
\psfrag{S0}[tc][tc]{\PFGstyle \footnotesize{$ 0$}}%
\psfrag{S10}[tc][tc]{\PFGstyle \footnotesize{$10$}}%
\psfrag{S2}[tc][tc]{\PFGstyle \footnotesize{$ 2$}}%
\psfrag{S4}[tc][tc]{\PFGstyle \footnotesize{$ 4$}}%
\psfrag{S6}[tc][tc]{\PFGstyle \footnotesize{$ 6$}}%
\psfrag{S8}[tc][tc]{\PFGstyle \footnotesize{$ 8$}}%
\psfrag{SuperscripA}[cc][cc]{\PFGstyle $\bar{E}_{\textsc{osc}}/N^{1/2}$ ($10^{-2}$)}%
\psfrag{Superscrip}[Bc][Bc]{\PFGstyle $N^{1/2}$}%
\psfrag{W00}[cr][cr]{\PFGstyle \footnotesize{$0.0$}}%
\psfrag{W05}[cr][cr]{\PFGstyle \footnotesize{$0.5$}}%
\psfrag{W10}[cr][cr]{\PFGstyle \footnotesize{$1.0$}}%
\psfrag{W15}[cr][cr]{\PFGstyle \footnotesize{$1.5$}}%
\psfrag{W20}[cr][cr]{\PFGstyle \footnotesize{$2.0$}}%
\psfrag{Wm05}[cr][cr]{\PFGstyle \footnotesize{$-0.5$}}%
\psfrag{Wm10}[cr][cr]{\PFGstyle \footnotesize{$-1.0$}}%
\psfrag{Wm15}[cr][cr]{\PFGstyle \footnotesize{$-1.5$}}%
\psfrag{Wm20}[cr][cr]{\PFGstyle \footnotesize{$-2.0$}}%
\centering
\includegraphics[width=68mm]{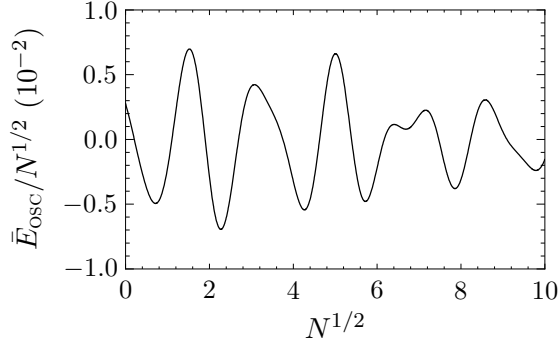}
\end{psfrags}\end{small}
\caption{Oscillating contribution to the ground state energy for the chaotic quantum dot of quartic confinement ($\lambda=1.2$), computed with the 30 shortest periodic orbits of the classical system.}
\label{E:OS}
\end{figure}

The oscillating corrections to the ground state energy are of order $N^\alpha$, and of period $N^{1/2}$, where $\alpha=3/4$ in the integrable  case and $\alpha=1/2$ in the fully chaotic case. Note that in both cases and contrary to the case of atoms, the oscillating corrections are less important than correlation effects. If $\V\FX$ is a harmonic well, the dynamics induced by $\W\FX$ is integrable and the oscillating part has been computed \cite{Plocek-2006}. It is particularly interesting to consider a case where the dynamics induced by $\W\FX$ is fully chaotic. For this purpose we have chosen the quartic potential $\V\FX=\epa(x_1^4/\epb+x_2^4\epb-\lambda x_1^2x_2^2)$, with $\epa=10^{-3}$ and $\epb=\pi/4$ \cite{BohigasTomsovicUllmo-PhysRep223,BrackEA-JPhysA36}. The nature of the dynamics of $\W\FX$ depends on the coupling constant $\epc$, and when $\lambda=1.2$, the dynamics is fully chaotic (whereas that of $\V\FX$ is not) \cite{Rueedi-2009}. In this case
\begin{equation}
\sEHFsm=
 0.14\ N^2
-0.067\ N^{3/2}
+0.053\ N
\end{equation}
and $\sEHFos$, computed by summing over the 30 shortest periodic orbits, is shown in figure \ref{E:OS}.

%
%
%
%
%
To conclude, we have found a systematic way to compute the ground state energy of atoms and quantum dots when the number of electrons is large. A semiclassical HF theory has been shown to give the dominant terms. But correlation effects have been computed which determine the validity of HF theory. We have also shown that shell effects result from the dynamics of a classical particle moving in the presence of the TF potential. This allows, in the case of dots, to see the impact on the oscillations of the energy of a regular or chaotic motion of this classical particle.

%
%
%
%
%
We thank C.~Plocek for useful and stimulating discussions at the early stage of this work. This work was supported by the Fonds National Suisse de la Recherche Scientifique.

%
%
%
%
%

\end{document}